\documentclass[11pt,twoside,a4paper]{article}
\usepackage{graphicx}
\usepackage[dvips]{epsfig}
\usepackage{amssymb}
\usepackage{rotating}
\usepackage{a4wide}

\newcommand{\be}{\begin{equation}}
\newcommand{\ee}{\end{equation}}
\newcommand{\ba}{\begin{eqnarray}}
\newcommand{\ea}{\end{eqnarray}}

\def\Journal#1#2#3#4{{#1} {\bf #2}, #3 (#4)}

\def\PLB{{\em Phys. Lett.}  B}
\def\PRL{\em Phys. Rev. Lett.}

\def\EPJC{{\em Eur. Phys.} J}

\def\Att{A_{TT}}
\def\sigup{\sigma^{\uparrow\uparrow}}
\def\sigdw{\sigma^{\uparrow\downarrow}}
\def\gsim{\mathrel{\rlap{\lower4pt\hbox{\hskip1pt$\sim$}}\raise1pt\hbox{$>$}}}

\begin{document}
\centerline{\Large \bf QCD physics with polarized antiprotons at GSI}
\vspace{1cm}

\begin{center}
  P. Lenisa$^1$, F. Rathmann$^2$, M. Anselmino$^3$, D.  Chiladze$^2$,
  M.  Contalbrigo$^1$, P.F. Dalpiaz$^1$, E. De Sanctis$^4$, A.
  Drago$^1$, A.  Kacharava$^5$, A.  Lehrach$^2$, B. Lorentz$^2$,
  G.~Macharashvili$^6$, R.  Maier$^2$, S. Martin$^2$, C.  Montag$^7$,
  N.N.  Nikolaev$^{2,8}$, E. Steffens$^5$, D.~Prasuhn$^2$, H.
  Str\"oher$^2$, and S.  Yaschenko$^5$
\end{center}

%\vspace{1cm}
\begin{center}
\it\small 
$^1$Universit\`a di Ferrara and INFN, 44100 Ferrara, Italy \newline 
$^2$Institut f\"ur Kernphysik, Forschungszentrum   J\"ulich, 52428 J\"ulich, Germany \newline
$^3$Dipartimento di  Fisica Teorica, Universit\`a di Torino and INFN, 10125 Torino,  Italy\newline
$^4$Istituto Nazionale di Fisica Nucleare, Laboratori  Nazionali di Frascati, 00044 Frascati, Italy\newline
$^5$Physikalisches Institut II, Universit\"at Erlangen--N\"urnberg,  91058 Erlangen, Germany\newline
$^6$Laboratory of Nuclear Problems,  Joint Institute for Nuclear Research, 141980 Dubna, Russia\newline
$^7$Brookhaven National Laboratory, Collider--Accelerator  Department, Upton, New York 11973--5000, USA\newline
$^8$L.D. Landau  Institute for Theoretical Physics, 142432 Chernogolovka, Russia
\end{center}

%\vspace{1cm}

\begin{abstract}
  A polarized antiproton beam at the Facility for Antiproton and Ion
  Research, proposed by the PAX collaboration, will open a window to
  new physics uniquely accessible at the new High Energy Storage Ring.
  Our proposal to realize an asymmetric collider, in which polarized
  protons with momenta of about 3.5~GeV/c collide with polarized
  antiprotons with momenta up to 15~GeV/c, is well--suited to perform
  a direct measurement of the transversity distribution function
  $h_1$.  In this report we summarize the outcome of various working
  group meetings within the PAX collaboration. The overall machine
  setup at the HESR, proposed by the PAX collaboration, is described
  along with the associated PAX experimental program.
\end{abstract}

The possibility to test the nucleon structure via double spin
asymmetries in polarized proton--antiproton reactions at the HESR ring
of FAIR at GSI has been suggested by the PAX collaboration last year
in Ref.~\cite{paxloi}.  Since then, there has been much progress, both
in understanding the physics potential of such an
experiment~\cite{jpsi,efremov,brodsky,zavada} and in studying the
feasibility of efficiently producing polarized antiprotons~\cite{ap}.
The physics program of such a facility would extend to a new domain
the exceptionally fruitful studies of the nucleon structure performed
in unpolarized and polarized deep inelastic scattering (DIS), which
have been at the center of high energy physics during the past four
decades. It suffices to mention the unique possibility of a direct
measurement of the transversity distribution function $h_1$, one of
the last missing fundamental pieces in the QCD description of the
nucleon. In the available kinematic domain of the proposed experiment,
which covers the valence region, the Drell--Yan double transverse spin
asymmetry was recently predicted to be as large as
30\%~\cite{jpsi,efremov}.  Other novel tests of QCD at such a facility
include the polarized elastic hard scattering of antiprotons on
protons and the measurement of the phases of the time--like form
factors of the proton (see \cite{paxloi}).  A viable practical
scheme\footnote{The basic approach to polarizing and storing
  antiprotons at HESR--FAIR is based on solid QED calculations of the
  spin transfer from electrons to antiprotons \cite{HOMeyer}, which is
  being routinely used at Jefferson Laboratory for the electromagnetic
  form factor separation \cite{JlabFF}, and which has been tested and
  confirmed experimentally in the FILTEX experiment \cite{Filtex}.}
which allows us to reach a polarization of the stored antiprotons at
HESR--FAIR of $\simeq 30\%$ has been worked out and published in Ref.~\cite{ap}.\\

%The PAX LoI was submitted on January 15, 2004. The physics program of
%PAX has been positively reviewed by the QCD--PAC on May 14--16, 2004.
%The proposal by the ASSIA collaboration \cite{ASSIA} to utilize a
%polarized solid target and bombard it with an unpolarized antiproton
%beam extracted from the synchrotron SIS100 (???) has been rejected by
%the GSI management. (???) Such measurements would not allow one to
%achieve the aspired goal anyway, because the transversity distribution
%function in single spin asymmetries appears always coupled to an
%unknown fragmentation function.  Following the QCD--PAC report and the
%recommendation of the Chairman of STI and the FAIR project
%coordinator, the PAX collaboration has optimized the technique to
%achieve a sizable antiproton polarization and the proposal for
%experiments at GSI with polarized antiprotons.

\begin{figure}[h]
\centerline{\includegraphics[width=0.80\linewidth]{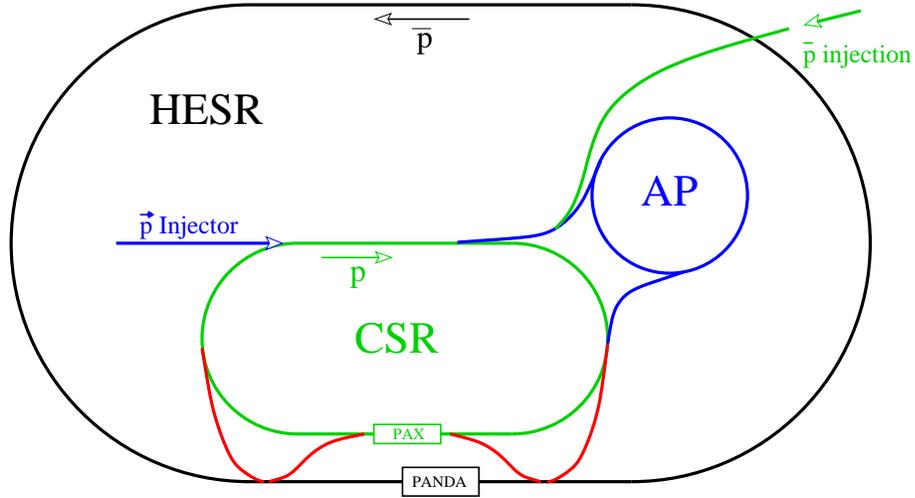}}
\caption{\small The proposed accelerator set--up at the HESR (black), with the 
  equipment used by the PAX collaboration in Phase I: CSR (green), AP,
  beam transfer lines and polarized proton injector (all blue). In
  Phase II, by adding two transfer lines (red), an asymmetric collider
  is set up. It should be noted that, in this phase, also fixed target
  operation at PAX is possible. }
\label{CSRring}
\end{figure}

In this brief report we summarize the outcome of various working group
meetings of the PAX Collaboration, the results of which have been
presented in 2004 at several workshops and conferences \cite{paxweb}.
%maybe we should quote here the PAX webpage, and in that webpage list
%also all the talks at conferences (Trento ECT*, SPIN2004, etc...)   
%This note summaries the outcome of the PAX work, anticipating
%the main lines of the emerging proposal, the physical motivations and the
%most important expected results. This might help your evaluation of the
%final complete project which will be presented in January.
%Still, achieving this ambitious goal requires crossing uncharted
%terrain and calls for the staged approach 
%We propose a staged approach in which all the
%major itemsf can be tested and optimized before going 
%to the most challenging part of the project: the polarized
%proton-antiproton collider. 
The PAX collaboration proposes an approach that is composed of two
phases. During these the major milestones of the project can be tested
and optimized before the final goal is approached: A polarized
proton--antiproton asymmetric collider, in which polarized protons
with momenta of about 3.5 GeV/c collide with polarized antiprotons
with momenta up to 15 GeV/c. These circulate in the HESR, which has
already been approved and will serve the PANDA experiment. In the
following, we will briefly describe the overall machine setup at the
HESR, schematically depicted in Fig.~\ref{CSRring}, as proposed by the
PAX collaboration.

Let us summarize the main features of the accelerator setup:
\begin{itemize}
\item[1.] An Antiproton Polarizer (AP) built inside the HESR area with
  the crucial goal of polarizing antiprotons at kinetic energies
  around $\approx 50$~MeV ($p\approx 300 $ MeV/c), to be accelerated
  and injected into the other rings.
  
\item[2.] A second Cooler Synchrotron Ring (CSR, COSY--like) in which
  protons or antiprotons can be stored with a momentum up to 3.5
  GeV/c.  This ring shall have a straight section, where a PAX
  detector could be installed, running parallel to the experimental
  straight section of HESR.
  
\item[3.] By deflection of the HESR beam into the straight section of
  the CSR, both the collider or the fixed--target mode become
  feasible.

\end{itemize}

It is worthwhile to stress that, through the employment of the CSR,
effectively a second interaction point is formed with minimum
interference with PANDA. The proposed solution opens the possibility
to run two different experiments at the same time.\\

The physics program should be pursued in two different phases.

\begin{itemize}
\item[(I)] A beam of unpolarized or polarized antiprotons with
  momentum up to 3.5 GeV/c in the CSR ring, colliding on a polarized
  hydrogen target in the PAX detector. This phase is independent of
  the HESR performance.
  
  This first phase, at moderately high energy, will allow for the
  first time the measurement of the time--like proton form factors in
  single and double polarized $\bar{p}p$ interactions in a wide
  kinematical range, from close to threshold up to $Q^2=8.5$~GeV$^2$.
  It would enable to determine several double spin asymmetries in
  elastic $\bar{p}^{\uparrow}p^{\uparrow}$ scattering.  By detecting
  back scattered antiprotons one can also explore hard scattering
  regions of large $t$: In proton--proton scattering the same region
  of $t$ requires twice the energy.  There are no competing facilities
  at which these topical issues can be addressed.  For the theoretical
  background, see the PAX LoI \cite{paxloi} and the recent review
  paper \cite{brodsky}.
  
\item[(II)] This phase will allow the first ever direct measurement of
  the quark transversity distribution $h_1$, by measuring the double
  transverse spin asymmetry $\Att$ in Drell--Yan processes
  $p^{\uparrow} \bar{p}^{\uparrow} \rightarrow e^+ e^- X$ as a
  function of Bjorken $x$ and $Q^2$ (= $M^2$)
  $$\Att \equiv \frac{d\sigup-d\sigdw}{d\sigup+d\sigdw}\,=\,
  \hat{a}_{TT}\frac{\sum_q e_q^2
    h_1^q(x_1,M^2)h_1^{\overline{q}}(x_2,M^2)} {\sum_q e_q^2
    q(x_1,M^2)\overline{q}(x_2,M^2)}\,,$$
  where
  $q=u,\overline{u},d,\overline{d}\ldots$, $M$ is the invariant mass
  of the lepton pair and $\hat{a}_{TT}$, of the order of one, is the
  calculable double--spin asymmetry of the QED elementary process
  $q\overline{q}\rightarrow e^+ e^-$.  Two possible scenarios might be
  foreseen to perform the measurement.
  \begin{itemize}
    
  \item[(a)] {\bf Asymmetric Collider:} A beam of polarized
    antiprotons from 1.5 GeV/c up to 15 GeV/c circulating in the HESR,
    colliding on a beam of polarized protons with momenta up to 3.5
    GeV/c circulating in the CSR.  This scenario however requires to
    demonstrate that a suitable luminosity is reachable.  Deflection
    of the HESR beam to the PAX detector in the CSR is necessary (see
    Fig.~\ref{CSRring}).
    
    By proper variation of the energy of the two colliding beams, this
    setup would allow a measurement of the transversity distribution
    $h_1$ in the valence region of $x>0.05$, with corresponding $Q^2=4
    \ldots 100$ $\rm GeV^2$ (see Fig.~\ref{Figphysics}). $\Att$ is
    predicted to be larger than 20 \% over the full kinematic range,
    up to the highest reachable center--of--mass energy of
    $\sqrt{s}\sim\sqrt{200}$. The cross section is large as well: With
    a luminosity of $5\cdot 10^{30}$~cm$^{-2}s^{-1}$ about $2000$
    events per day can be expected\footnote{A first estimate indicates
      that in the collider mode luminosities in excess of
      $10^{30}$~cm$^{-2}$s${^-1}$ could be reached. We are presently
      evaluating the influence of intra-beam scattering, which seems
      to be one of the limiting factors.}. For the transversity
    distribution $h_1$, such an experiment can be considered as the
    analogue of polarized DIS for the determination of the helicity
    structure function $g_1$, i.e. of the helicity distribution
    $\Delta q(x,Q^2)$; the kinematical coverage $(x,Q^2)$ will be
    similar to that of the HERMES experiment.
    
  \item[(b)] {\bf High luminosity fixed target experiment:} If the
    required luminosity in the collider mode is not achievable, a
    fixed target experiment can be conducted.  A beam of 22 GeV/c (15
    GeV/c) polarized antiprotons circulating in the HESR is used to
    collide with a polarized internal hydrogen target.  Also this
    scenario requires the deflection of the HESR beam to the PAX
    detector in the CSR (see Fig.~\ref{CSRring}).
  
    A theoretical discussion of the significance of the measurement of
    $\Att$ for a 22~GeV/c (15~GeV/c) beam impinging on a fixed target
    is given in Refs. \cite{jpsi,efremov,zavada} and the recent review
    paper \cite{brodsky}. The theoretical work on the $K$--factors for
    the transversity determination is in progress
    \cite{Ratcliffe,barone}. This measurement will explore the valence
    region of $x>0.2$, with corresponding $Q^2=4 \ldots 16$ ${\rm
      GeV}^2$ (see Fig.~\ref{Figphysics}).  In this region $\Att$ is
    predicted to be large (of the order of 30 \%, or more) and the
    expected number of events can be of the order of 2000 per day.
%This kinematic region
%  results to be complementary to the one covered in phase III: when
%  the Bjorken $x$ of the proton and $\bar{x}$ of the antiproton
%  coincide, $\Att(x,Q^2)$ is simply proportional to $[h_1(x,Q^2)]^{2}$
%   This measurement will allow us to extract unambiguously for the
%   first time the valence quark transverse spin structure function.
%Detailed evaluations can be found in the PAX LoI \cite{paxloi}.
\end{itemize}

%Such a measurement of $h_1$ at FAIR will accomplish a task analogous to
%that of polarized DIS for the helicity distribution Delta q.

\begin{figure}[hbt]
  \includegraphics[width=0.50\linewidth]{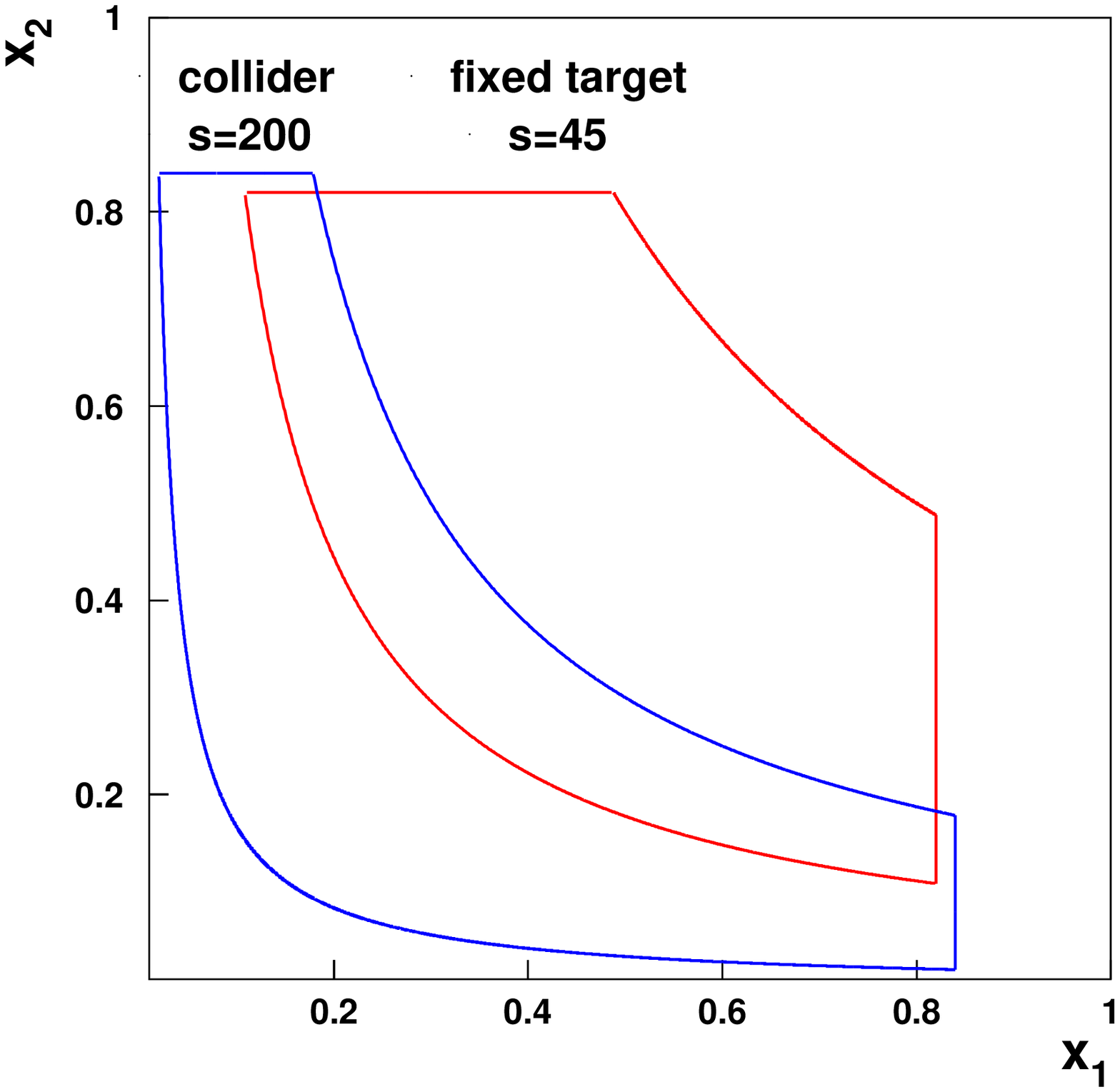}
  \includegraphics[width=0.50\linewidth]{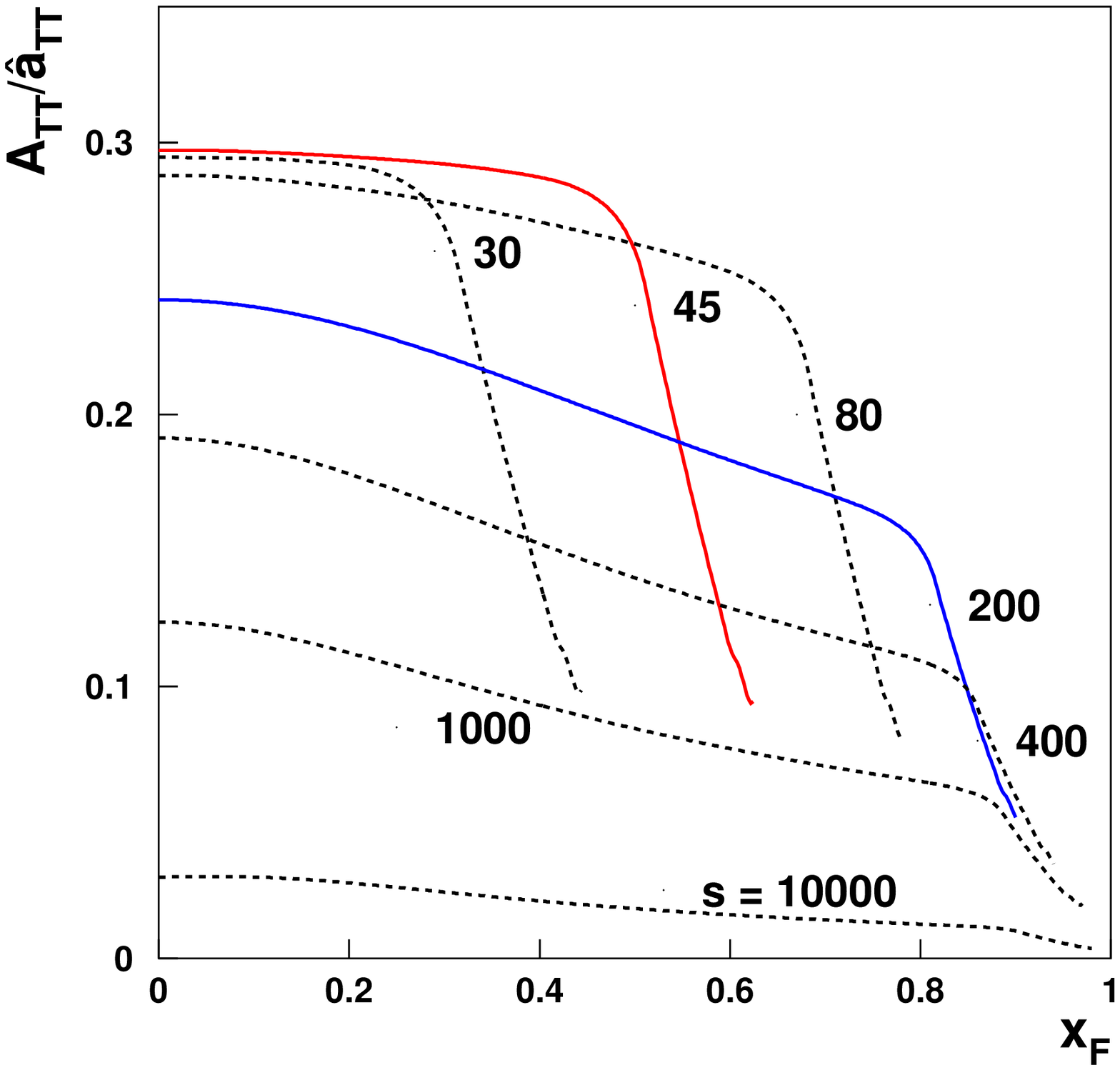}
\caption{\small Left: The kinematic region covered by the $h_1$ measurement 
  at PAX in phase II. In the asymmetric collider scenario (blue)
  antiprotons of 15~GeV/c impinge on protons of 3.5~GeV/c at c.m.
  energies of $\sqrt{s}\sim \sqrt{200}$~GeV and $Q^2>4$ $\rm GeV^2$.  
  The fixed target case
  (red) represents antiprotons of 22~GeV/c colliding with a fixed
  polarized target ($\sqrt{s}\sim\sqrt{45}$~GeV).  Right: The expected
  asymmetry as a function of Feynman $x_F$ for different values of
  $s$ and $Q^2=16$ $\rm GeV^2$.}
\label{Figphysics}
\end{figure}

We would like to mention, that we are also investigating whether the
PANDA detector, properly modified, is compatible with the transversity
measurements in the collider mode, where an efficient identification
of the Drell--Yan pairs is required. At the interaction point, the
spins of the colliding protons and antiprotons should be vertical,
with no significant component along the beam direction.

\end{itemize}
%Kolya2
%{\it Here must go a new figure from Drago.
%We just notice that the range of $x,Q^2$ accessible at such a collider
%will be comparable to, and aven broader, than that explored in the
%HERMES experiment at HERA. }

%Kolya2

%The physics with polarized antiprotons opens up many unique possibilities:
%that was discussed in the PAX LoI [1] and an independent recent review paper
%can be found in Ref. [2]. The technique to build up (in the AP) a sizable
%antiproton polarization has been throughly discussed, is now well
%established, and is presented in Ref. [3]. 
%We mention in passing that
%the electron-to-proton polarization transfer is actively been used at
%Jefferson Lab for separation of the charge and magnetic form factors of the
%proton and neutron [3*]. 

%We recall here only the main new
%physical results expected from the three phases.

To summarize, we note that the storage of polarized antiprotons at
HESR will open unique possibilities to test QCD in hitherto unexplored
domains.  This will provide another cornerstone to the antiproton
program at FAIR.

\end{document}